\documentclass[10pt]{article}

\usepackage{amsmath}

\begin{document}


\begin{center}
{\Large {\bf Schmidt states and positivity of linear maps }}
\end{center}

\hfil\break

\begin{center}
{\bf Aik-meng Kuah\footnote{kuah@physics.utexas.edu}, E.C.G. Sudarshan}\break
{\it Department of Physics \\
University of Texas at Austin \\
Austin, Texas 78712-1081}\\
\end{center}

\begin{center}
Jun 12 2005
\end{center}

\begin{abstract}

Using pure entangled Schmidt states, we show that m-positivity of a map is bounded by the ranks of its negative Kraus matrices.  We also give an algebraic condition for a map to be m-positive.  We interpret these results in the context of positive maps as entanglement witnesses, and find that only 1-positive maps are needed for testing entanglement.

\end{abstract}

\hfil\break
\hfil\break

\pagebreak


\section{Introduction}

A linear map of $N$ dimensional density matrices is itself a matrix (or super matrix) of $N^2$ dimensions, and the matrix indices can be arranged in several ways.  For our paper we will write maps in what we call the B-form, where a map $\Lambda$ acting on a density matrix $\rho$ is given by:

\begin{equation}
[\Lambda(\rho)]_{i'j'} = \Lambda_{i'i,j'j} \rho_{ij}
\end{equation}

It was shown in \cite{Sudarshan61} that when written in B-form, the map matrix is hermitian, and if trace preservation and complete positivity is required, the map can be seen as a reduced unitary transformation.

In general physical dynamical maps are completely positive (although in certain specific problems on a restricted domain, the actual dynamics of a system may be given by non-positive maps \cite{JordanShajiSudarshan04}).  However positive, but not completely positive, maps are important to the problem of understanding entanglement.

A map $\Lambda$ is defined as m-positive if the map $\Lambda \otimes I_m$ is positive.  This notion of m-positivity is a little perplexing, since both $I_m$ is clearly positive, and $\Lambda$ is positive on its original domain.  

It was then shown in \cite{Horodecki96} that positive but not completely positive maps form a set of linear functionals that delineate the convex set of separable states from entangled states.  It is clear that the map $\Lambda \otimes I_m$ would have to be positive on separable states, so if it is not m-positive then it must be because it is not positive on an entangled state.

Understanding the positivity problem could bring us closer to a simple test for entanglement – for example in \cite{Horodecki96} where a property of low dimensional maps due to Stormer \cite{Stormer63} and Woronowicz \cite{Woronowicz76} was used to prove the famous Peres partial transpose condition \cite{Peres96} for separability in the 2 dimensional case.

In this paper we show that positivity of a map depend on the ranks of its Kraus matrices.  We provide a upper bound condition on the positivity of a given map, and give an algebraic method of checking if a map is m-positive.  We consider the example of N=2 1-positive maps.  Lastly we interpret our results in the context of entanglement witnesses.

\section{An upper bound on positivity}

A map $\Lambda$ written in B-form is hermitian, so it can be decomposed in canonical form \cite{Sudarshan61}:

\begin{equation}
\Lambda_{i'i,j'j} = \sum_\alpha^{N^2} \lambda^\alpha L^\alpha_{i'i} {L^\alpha_{jj'}}^\dagger 
\end{equation}

Where the eigen-matrices (also known as Kraus matrices) $L^\alpha$ are orthornormal (note that we did not absorb the eigen-values $\lambda^\alpha$ into the definition of the eigen-matrices $\Lambda^\alpha$ as was done in \cite{Sudarshan61}):

\begin{equation}
\sum_{i'i} L^\alpha_{i'i} {L^\beta_{i'i}}^* = \delta_{\alpha,\beta}
\end{equation}

If $\Lambda$ has a negative eigen-value $\lambda^Q$ and corresponding Kraus matrix $L^Q$, then we can show that $\Lambda$ cannot be $rank(L^Q)$-positive.

Take the polar decomposition of $L^Q = U^Q M^Q$ where $U^Q \in SU(N)$ and $M^Q$ is hermitian matrix.  Let $|i^Q\rangle$ be the eigen-vectors of $M^Q$, and let us define the Schmidt entangled state:

\begin{equation}
|\Phi\rangle = \frac{1}{\sqrt{r}} \sum_i^{r} |i^Q\rangle \otimes |i\rangle 
\end{equation}

Where $r \equiv rank(L^Q)=rank(M^Q)$.  Consider the action of the map $\Lambda \otimes I_r$ acting on this Schmidt state, we find that the resulting density matrix $\Lambda \otimes I (|\Phi\rangle\langle\Phi|)$ is actually just a $rN \times rN$ sub-matrix of $\Lambda$ in B-form:

\begin{equation}
\Lambda \otimes I (|\Phi\rangle\langle\Phi|)
= \frac{1}{r} \sum_{ij}^r \Lambda(|i^Q\rangle\langle j^Q|)\otimes |i\rangle\langle j|
= \frac{1}{r} \sum_{i'j'}^N \sum_{ij}^r \Lambda_{i'i^Q,j'j^Q} |i'i\rangle\langle j'j|
\end{equation}

Since we used the eigen-vectors of $M^Q$, this sub-matrix must contain a full copy of the Kraus matrix $L^Q$.  And because the Kraus matrices are orthornormal, this $rN \times rN$ sub-matrix must itself have an eigen-decomposition that includes the eigen-value $\lambda^Q$ and eigen-matrix $L^Q$.  Therefore $\Lambda \otimes I_r (|\Phi\rangle\langle\Phi|)$ is not positive, so $\Lambda$ is not $rank(L^Q)$-positive.

Note that for our entangled test state, we had used the eigen-vectors $|i^Q\rangle$, but any set of orthornormal vectors with the same span will also do.  This gives us a side result -- if $rank(L^Q)=N$, any Schmidt rank N state will have the same (complete) span, so the map $\Lambda \otimes I$ will be negative on all Schmidt rank N states regardless of the basis.

The ranks of the negative Kraus matrices give an upper bound on the positivity of the map.  However, the map could still be non-positive in smaller dimensions, we will discuss this problem in a later section.

\section{N-positivity is complete positivity}

In this section we provide an alternative proof of Choi's theorem \cite{Choi72} -- an N-positive map is completely positive.

For the maximally entangled Schmidt state $|\Phi\rangle = 1/\sqrt{N} \sum_i^N |i\rangle \otimes |i\rangle$, the density matrix $\Lambda \otimes I_N (|\Phi\rangle\langle\Phi|)$ is simply given by the B-form matrix $\Lambda$ itself.  The eigen-values and Kraus matrices of the map are correspondingly the eigen-values and eigen-vectors $|\epsilon^\alpha\rangle$ of the density matrix $\Lambda \otimes I (|\Phi\rangle\langle\Phi|)$:

\begin{equation}
|\epsilon^\alpha\rangle \equiv \sum_{i'i}^N L^\alpha_{i'i} |i'\rangle \otimes |i\rangle
\end{equation}

Therefore it is sometimes convenient to think of the map itself as a pseudo density matrix, and the Kraus matrices as eigen-vectors of the pseudo density matrix.

It is clear that the map is $N$-positive iff it is positive on this maximally entangled Schmidt state.  Now consider the positivity of $\Lambda \otimes I_m$ where $m > N$.  Any pure state of this space has a maximum Schmidt decomposition of $N$ terms, therefore $\Lambda$ is $m$-positive if it is positive on all Schmidt rank $N$ pure states, ie. if $\Lambda$ is $N$-positive.  Therefore N-positivity is necessary and sufficient for complete positivity.

\section{General condition for m-positivity}

The necessary and sufficient condition that the map is m-positive is that it must be positive on all Schmidt rank m pure states.  A Schmidt rank m state can be written:

\begin{equation}
|\Phi\rangle = \sum_i^{m} s_i |i\rangle \otimes |i\rangle
\end{equation}

The map $\Lambda \otimes I$ acting on this entangled state gives the sub-matrix:

\begin{equation}
s_i s_j^* \Lambda_{i'i,j'j} 
\end{equation}

The Schmidt coefficients $s_i$ do not alter the positivity of this matrix -- if for some $x^\dagger_{i'i} \Lambda_{i'i,j'j} x_{j'j} < 0$, there exists $y_{i'i} = s_i^{-1} x_{i'i}$ which makes $y^\dagger_{i'i} (s_i s_j^* \Lambda_{i'i,j'j}) y_{j'j} < 0$.  Therefore the result is similarly positive or negative for all states with the same Schmidt basis, regardless of the Schmidt coefficients.  So to check if a map is m-positive, we can use Schmidt rank m states with degenerate Schmidt coefficients, but over all possible Schmidt basis:

\begin{equation}
|\Phi\rangle = \frac{1}{\sqrt{m}} \sum_i^{m} \sum_{jk}^N U_{ji} |j\rangle \otimes V_{ki} |k\rangle
\end{equation}

Where $U,V \in SU(N)$.  The state gets mapped to:

\begin{equation}
\Lambda \otimes I (|\Phi\rangle\langle\Phi|) = \frac{1}{m} \sum_{j',j,l',l}^N  \Lambda_{j'j,l'l} \sum_i^m(U_{ji} V_{ki}) \sum_h^m(U_{lh}^* V_{oh}^*) |j'\otimes k\rangle\langle l' \otimes o| 
\end{equation}

Let us define a m-dimensional projection $P \equiv \sum_i^m |i\rangle\langle i|$, so we can write the condition for positivity as:

\begin{equation}
V_{ki} P_i U^T_{ij} \Lambda_{j'j,l'l} U^*_{lh} P_h V^\dagger_{ho} \geq 0
\end{equation}

We can see that $V$ is unnecessary and the condition can be simplified to:

\begin{equation}
(I \otimes (P U)) \Lambda (I \otimes (P U)^\dagger) \geq 0
\end{equation}

Where $U \in SU(N)$ .  Finally we note that it is possible to further restrict $U$; since only the part of the map containing the negative Kraus matrix is important, we need only check for $U$ where the projected submatrix contains part or all of at least one negative Kraus matrix.

\section{N=2 1-positive maps}

Let us apply the results from the previous sections to the analysis of N=2 positive maps.  We now know that for N=2, if the map is 1-positive the rank of the negative Kraus matrices must be 2.  Equivalently, if we treat the map matrix as a pseudo density matrix, then the negative eigen-vectors must have Schmidt rank 2.

We can show that a 1-positive map for N=2 can have only 1 negative eigen-vector by process of elimination.  First let us consider the case if the map has 3 negative eigen-vectors.  Since the negative eigen-vectors must be Schmidt rank 2, we can write eigen-values and eigen-vectors in general as:

\begin{eqnarray}
A, a|00\rangle + b|11\rangle \\
B, b^*|00\rangle - a^*|11\rangle \\
C, c|01\rangle + d|10\rangle \\
D, d^*|01\rangle – c^*|10\rangle
\end{eqnarray}

The density matrix has the form:

\begin{equation}
\left(\begin{array}{cccc}
Aa^2 + Bb^2 & 0 & 0 & (A-B)ab^* \\
0 & Cc^2 + Dd^2 & (C-D)cd^* & 0 \\
0 & (C-D)c^*d & Cd^2 + Dc^2 & 0 \\
(A-B)a^*b & 0 & 0 & Ab^2 + Ba^2
\end{array}\right)
\end{equation}

From the last section, we use the projection $P = I \otimes |0\rangle\langle 0|$ and $U = 1$ and obtain two necessary conditions for 1-positivity:

\begin{eqnarray}
Aa^2 + Bb^2 \geq 0 \\
Cc^2 + Dd^2 \geq 0
\end{eqnarray}

We can see that if any 3 of the eigen-values A, B, C or D are negative, the map cannot be 1-positive.

Next let us consider if 2 eigen-values can be negative.  From the above necessary condition for positivity, we know that no pair of coefficients A,B or C,D can be negative.  The only possibility is if one coefficient of each pair A,B and C,D is negative.  Using the same density matrix above and a generic unitary transformation:

\begin{eqnarray}
U |0\rangle = \alpha |0\rangle + \beta |1\rangle \\
U |1\rangle = \beta^* |0\rangle - \alpha^* |1\rangle
\end{eqnarray}

The projected matrix $(I \otimes PU) \Lambda (I\otimes PU)^\dagger$ is:

\begin{equation}
\left(\begin{array}{cc}
(Aa^2 + Bb^2)\alpha^2 & (A-B)ab^*\alpha\beta \\
(A-B)a^*b\alpha^*\beta^* & (Ab^2 + Ba^2)\beta^2 \\
\end{array}\right)
+ 
\left(\begin{array}{cc}
(Cc^2 + Dd^2)\beta^2 & (C-D)cd^*\alpha^*\beta^* \\
(C-D)c^*d\alpha\beta & (Cd^2 + Dc^2)\alpha^2 \\
\end{array}\right)
\end{equation}

A quick calculation of the determinant of each matrix tells us each matrix is negative.  For the sum of 2 negative matrices to be positive, the positive eigen-vector of one matrix must cancel the negative eigen-vector of the other.  We notice that the unitary transformation $U$ transforms the 2 matrices differently, therefore the matrices cannot balance for all $U$ unless the matrices are invariant under $U$ (ie. proportional to I), which would contradict the requirement that 2 eigen-values are negative.  Therefore we conclude that for N=2, a 1-positive map can have only 1 negative eigen-value.

\section{Transposition map}

The simplest and most well known example of a N=2 1-positive map is the transposition map:

\begin{equation}
T(\rho) = I_2 - \rho
\end{equation}

The transposition map has three Kraus matrices $\sigma_x$, $\sigma_y$, $\sigma_z$ with positive eigenvalues 1/2 and one Kraus matrix $I$ with negative eigenvalue -1/2.  

This negative Kraus matrix is rank 2 so the map $T \otimes I$ would be negative on all Schmidt rank 2 pure states.  Therefore for N=2, the transposition map alone is sufficient to differentiate all entangled pure states from separable pure states.

\section{Positive maps as entanglement witnesses}

The importance of positive but not completely positive maps as entanglement witnesses was given in \cite{Horodecki96}.  The set of all positive maps form a set of linear functionals separating the convex set of separable states from entangled states.

From the example of the transposition map, we now have a better picture of this geometry.  Consider a map that has a negative Kraus matrix of rank m+1, and having suitable positive Kraus matrices so that it is m-positive.  This map will be negative on some pure states of Schmidt rank m+1 and greater, but will be positive on all pure states of Schmidt rank m and lower.  Therefore the set of all such m-positive maps separate pure states of Schmidt rank $\geq m+1$ and pure states of Schmidt rank $\leq m$.   

For the purpose of an entanglement witness, it is only necessary to use the set of 1-positive maps.  Although we do not need all m-positive maps for determining entanglement, they can be considered to separate various entanglement "classes", given by the Schmidt rank of the entangled state.  Note that for rank=N class, only one N-positive map is needed for the test.

\section{Conclusions}

We find that the m-positivity of maps is closely related to its action on Schmidt rank m entangled states.  With some of the results here, it may be possible to obtain a set of rules for constructing an arbitrary m-positive map.  We hope this will bring us closer to a simple method for testing entanglement.


\end{document}